\documentclass[twocolumn,epjc3]{svjour3}  
\journalname{Eur. Phys. J. C}
\usepackage{amsmath,amssymb,graphics,graphicx,multirow,makecell}
\usepackage[style=nature, backend=biber]{biblatex}
\usepackage{subfig}
\usepackage{float}
\usepackage{lipsum}
\usepackage[british]{babel}

\let\oldhat\hat

\renewcommand{\hat}[1]{\oldhat{\mathbf{#1}}}

\newcommand{\onbb}{$0\nu\beta\beta$}

\newcommand{\ogee}{$e^+e^-$}
\newcommand{\gerda}{\textsc{Gerda}}
\newcommand{\legend}{\textsc{Legend}}

\newcommand{\rn}{$R_{90}$}

\newcommand{\Th}{$^{228}$Th}
\newcommand{\qbb}{$Q_{\beta\beta}$}
\newcommand{\cofs}{$^{56}$Co}

\graphicspath{{figures/}}

\usepackage{lineno}
\begin{document}

\title{Topologies of $^{76}$Ge double-beta decay events and calibration procedure biases
}


\author{Tommaso Comellato\thanksref{addr1, e1}
        \and
        Matteo Agostini\thanksref{addr2, addr1, e2}
        \and
        Stefan Schönert\thanksref{addr1, e3}
}

\thankstext{e1}{e-mail: tommaso.comellato@tum.de}
\thankstext{e2}{e-mail: matteo.agostini@ucl.ac.uk}
\thankstext{e3}{e-mail: schoenert@ph.tum.de}



\institute{
        Physik Department E15, Technische Universität München, James-Franck-Straße 1, 85748, Garching, Germany \label{addr1}
        \and
        Department of Physics and Astronomy, University College London, Gower Street, London WC1E 6BT, UK \label{addr2}
}

\date{Received: date / Accepted: date}

\maketitle

\begin{abstract}
The analysis of the time profile of electrical signals produced by energy depositions in germanium detectors allows discrimination of events with different topologies. This is especially relevant for experiments searching for the neutrinoless double beta decay of $^{76}$Ge to distinguish the sought-after signal from other background sources.
The standard cali\-bration procedures used to tune the selection criteria for double-beta decay events use a \Th\ source, because it provides samples of signal-like events. These samples exhibit energy spatial distributions with subtle different topologies compared to neutrinoless double-beta decay events. In this work, we will characterize these topological differences and, with the support of a \cofs\ source, evaluate biases and precision of calibration techniques which use such event samples. Our results will be particularly relevant for future experiments in which a solid estimation of the efficiency is required.

\end{abstract}


\section{Introduction}
\label{intro}

The search for neutrinoless double-beta decay (\onbb) is one of the hottest topics in particle physics. Its discovery would unambiguously prove that neutrinos are Majorana particles, i.e. they are their own antimatter counterpart, and lepton number is not a global symmetry of the Standard Model \cite{Agostini.2022}.  
\onbb\ is not only predicted by our leading theories of why neutrinos are massive particles, but also by those explaining the matter\--anti\-matter asymmetry of our universe~\cite{rodejohann:neutrinos, deppisch:bsm, davidson:leptogenesis}.

\onbb\ is a nuclear transition in which two neutrons decay simultaneously into two protons and two electrons, preserving the baryon number but changing the lepton number by two units. Several isotopes have been used to search for such transition, each offering different detection techniques. Historically, $^{76}$Ge has always provided among the most stringent constraints on the half-life of the process. Searches for \onbb\ of $^{76}$Ge are carried out with high-purity germanium detectors built from material isotopically enriched up to 92\% in $^{76}$Ge. 
Such a detection concept offers several advantages:
Ge detectors have negligible radioactive internal contaminations~\cite{gerda:uraniumth}, a per-mill energy resolution, and advanced event reconstruction capabilities~\cite{gerda.PSD.2022}.
In addition, the high density of germanium crystals ensures that the two electrons emitted in \onbb\ are absorbed within a few millimeters from the decay vertex, generating well localized energy depositions which are fully contained within the detector~\cite{fiorini:0nbb}.
Ge detectors are also a well consolidated technology, broadly used for gamma-ray spectroscopy and radioactivity monitoring, and can be reliably produced in collaboration with industrial partners. 

\onbb\ events in a Ge detector result in mono-energetic and well-localized energy depositions. The whole decay energy is transferred to the two emitted electrons, whose summed kinetic energy is equal to the $^{76}$Ge Q-value, i.e. \qbb$=$2039\,keV. The two electrons are likely to share evenly the decay energy, but extreme cases in which one electron takes most of the energy are also possible. Electrons at these energies have an absorption length of about a millimeter in Ge. However, secondary Bremsstrahlung photons produced during the electron absorption can occasionally travel several millimeters from their production vertex before interacting, producing secondary energy deposition sites. Thus, the topology of \onbb\ events is complex and its understanding is fundamental to develop techniques for discriminating the sought-after signal from backgrounds such as gamma-rays scattering multiple times within the detector. 

In this manuscript, we first characterize the energy spatial distribution of \onbb\ and other classes of events occurring in Ge detectors. As all events result in the generation of primary electrons/positrons, we first study the energy deposition of single-electron e\-vents as a function of the electron kinetic energy (Sec.~\ref{subsec:absorption_electron}), and then evaluate the difference between them and more complicated classes of events (Sec.~\ref{subsec:onbb-like_events}). In Sec. ~\ref{sec:aoe} we establish a connection between the event spatial distributions and the estimator used in \onbb\ experiments to discriminate signal-like events from background. In Sec.~\ref{subsec:calibration} we first review how such estimator is typically calibrated, and then evaluate the biases and precision of such calibration procedures on the \onbb-tagging efficiency. Finally, in Sec.~\ref{sec:energy-dependent-tagging}, a \cofs\ source will allow for an experimental determination of these biases.

\section{Characterization of the event spatial distribution}
\label{sec:r90}

\subsection{Absorption of electrons in Ge}
\label{subsec:absorption_electron}

\begin{figure}[ht]
  \centering
  \subfloat[width=0.47\textwidth][]{\includegraphics[width=.48\textwidth]{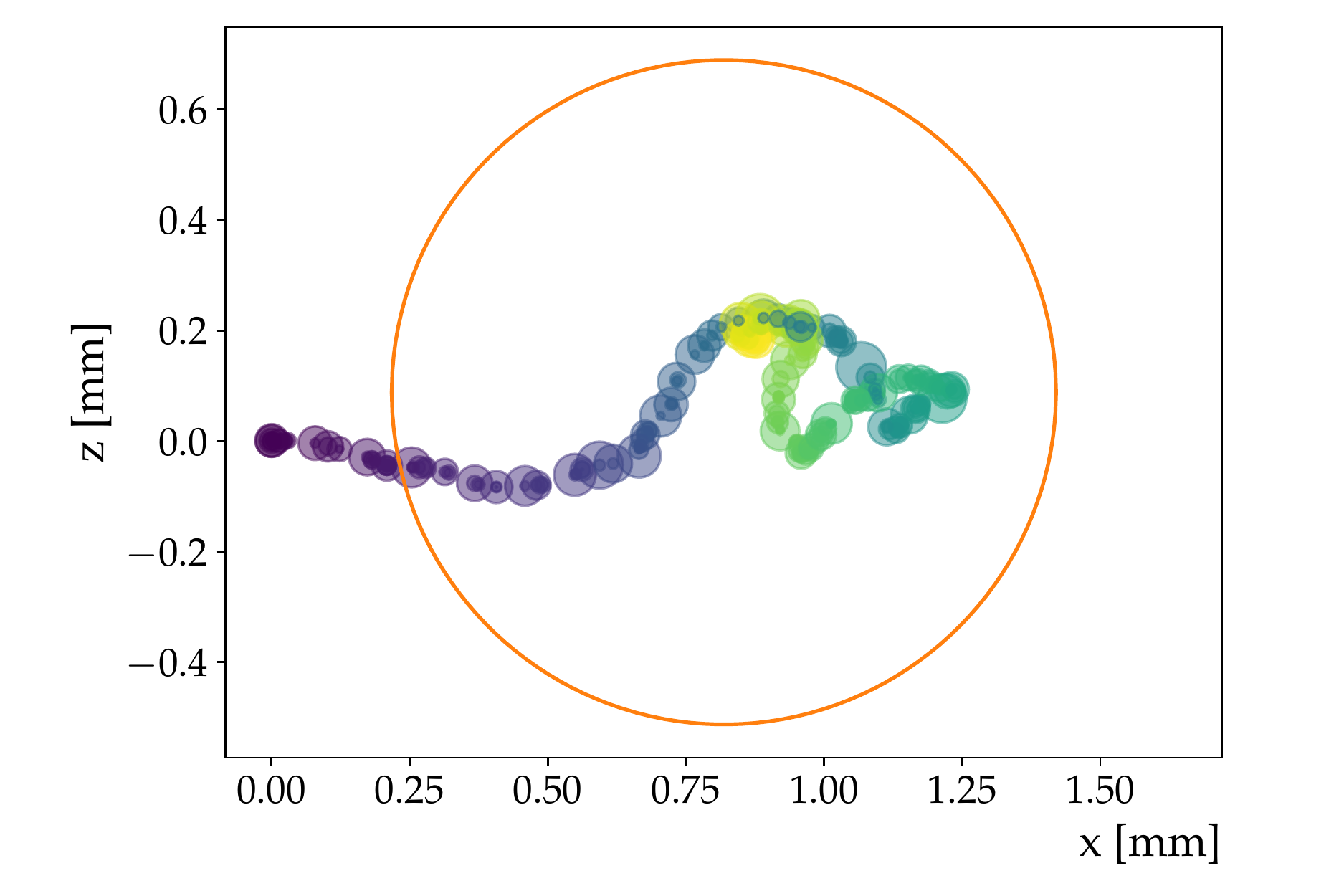}\label{subfig:electronWalk_noGamma}}\\

  \subfloat[width=0.47\textwidth][]{\includegraphics[width=.48\textwidth]{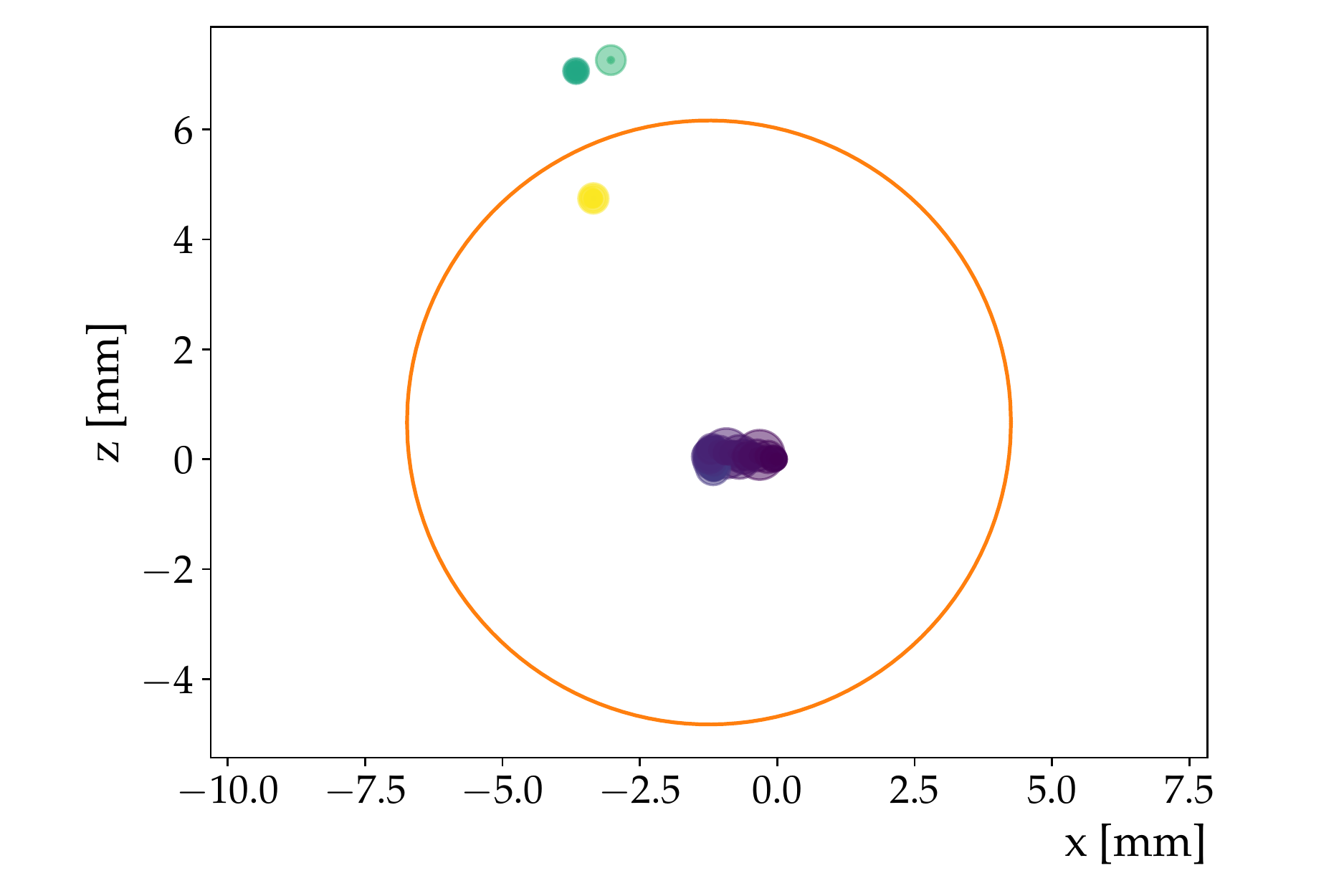}\label{subfig:electronWalk_wGamma}}\\

  \caption{Energy depositions of a $2$ MeV electron in germanium, through collisional \ref{sub@subfig:electronWalk_noGamma} and collisional plus radiative \ref{sub@subfig:electronWalk_wGamma} losses. The color code represents the time of each energy deposition, which flows from blue to yellow. The orange circle is centered around the center of energy and its radius is the \rn\ parameter described in the text. The path was simulated in 3 dimensions, and is reported here in cylindrical coordinates with respect to the starting point, which was taken as origin.}
  \label{fig:electronWalk}       
\end{figure}

%
\begin{figure*}[ht]
  \centering
  \subfloat[][\rn\ distribution from a primary electron with an energy of 2 MeV. The line histogram shows the total distribution, and the filled ones represent the subsets of events where an energy $E_\gamma$ higher than 50 and 200 keV is converted into Bremsstrahlung radiation.]{\includegraphics[width=.48\textwidth]{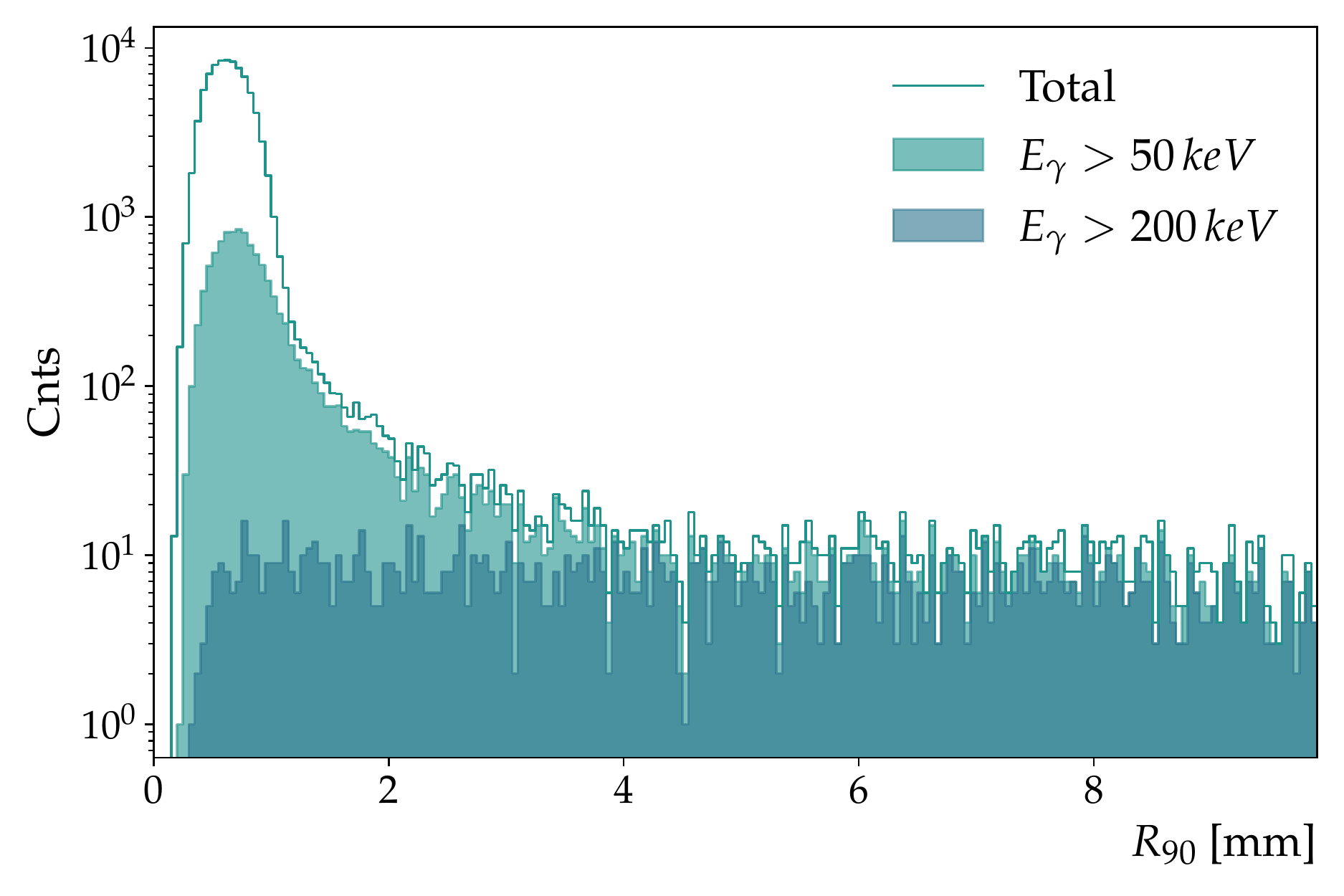}\label{subfig:hr90}} \quad
  \subfloat[][Dependence of the \rn\ distributions on energy. The green line follows the dependence of the peak, the yellow, light and dark green bands the 85\%, 90\% and 95\% quantiles, respectively]{\includegraphics[width=.48\textwidth]{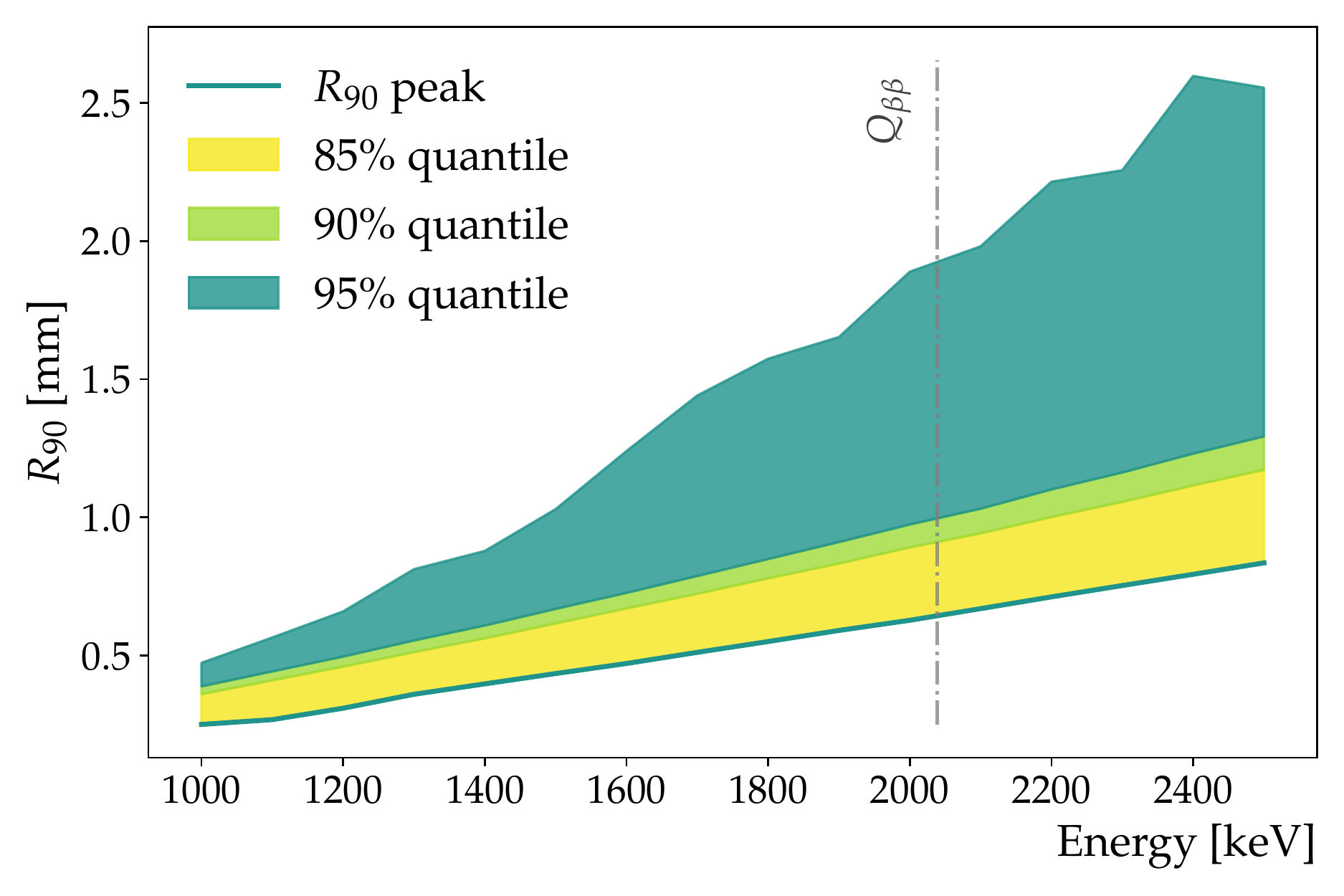}\label{subfig:R90vsE}} \\

  \caption{\rn\ distributions from absorption of monoenergetic electrons in germanium, and their dependence on energy.}
  \label{fig:hr90}       
\end{figure*}

Electrons lose energy mainly through collisional losses, i.e. ionization and excitation. As the energy loss occurs via interaction with orbiting electrons, a large fraction can be lost in a single collision. For this reason, the electron trajectory follows a tortuous path, where the energy loss per collision is inversely proportional to the electron kinetic energy. An example of path for an electron depositing $2$ MeV through collisions in Ge is shown in Fig.~\ref{subfig:electronWalk_noGamma}. Every point in the plot represents a collision, in which an energy proportional to the area of the marker is transferred to the medium. The color code marks the time of each energy deposition, which flows from blue to yellow, with a time scale which is of the order of a few ps. The electron in Fig.~\ref{subfig:electronWalk_noGamma} moves from left to right almost undisturbed, until its energy is comparable to that of the orbiting electrons. This occurs after about $1$ mm, when the electron begins a Brownian motion.

Electrons can also lose energy radiating Brems\-strah\-lung gammas, which occasionally travel up to few centimeters in Ge, producing secondary interaction sites. This is shown in Fig.~\ref{subfig:electronWalk_wGamma}, where a gamma-ray with an energy of about $200$ keV travels several millimeters upwards before undergoing multiple Compton scatterings. 

Therefore, according to the case, energy depositions from electrons with the same energy can produce very different spatial distributions. To characterize them we use the \rn\ parameter. This is defined as the minimum radius of the sphere which is centered in the energy weighted average of all energy depositions and contains $90\%$ of the total deposited energy. As shown in Fig.~\ref{fig:electronWalk}, in the case of energy loss through collisions (Fig.~\ref{subfig:electronWalk_noGamma}) this parameter is smaller than $1$ mm, while it can increase by an order of magnitude if a gamma-ray with sufficient energy is also emitted (Fig.~\ref{subfig:electronWalk_wGamma}). 

We can thus use the \rn\ parameter to study the size of the spatial distributions of electrons depositing energy in germanium and estimate both qualitatively and quantitatively how they are affected by the emission of Bremsstrahlung gammas. To this purpose, we ran a set of Monte Carlo simulations using the MaGe software framework \cite{MaGe} and the Geant4 toolkit \cite{geant4}. The simulated setup consisted of an inverted coaxial Ge detector, which is the state-of-the-art geometry developed for the future \onbb-experiment \textsc{Legend} \cite{Legend.pCDR}. The detector specifications and a validation of the simulation are discussed in Ref.~\cite{comellato:modelingIC}.

Fig.~\ref{subfig:hr90} shows the \rn\ distribution for single-electron events with an initial kinetic energy of $2$ MeV. The distribution is characterized by a Gaussian peak at $0.6$ mm and a tail of events extending to higher values. The events in the tail are those in which Bremsstrahlung gammas are produced, as shown by the \rn\ distribution of the subsets of events in which a gamma-ray with energy higher than $50$ keV or $200$ keV is emitted. 

The event topology is affected by the electron initial kinetic energy, both in the collisional and in the radiative sector. The centroid of the gaussian peak and the quantiles of the distributions as a function of the kinetic energy are shown in Fig.~\ref{subfig:R90vsE}.
As the scattering length of electrons increases with energy, the spatial distributions get on average broader. This is captured by the position of the \rn\ peak, which shifts from a value of $0.2$ mm at $1$ MeV, to $0.8$ mm at $2.5$ MeV. Naturally, the variance $\sigma_{R_{90}}$ of the peak also increases, as reported in Tab.~\ref{tab:r90_char}.
The higher the energy of the electron, the larger is the probability of a high energetic gamma-ray emission. As a consequence, the quantiles of the distributions increase with different dependences, with the tail extending to larger and larger \rn\ values. 

\begin{figure}[ht]
  \centering
  \includegraphics[width=0.48\textwidth]{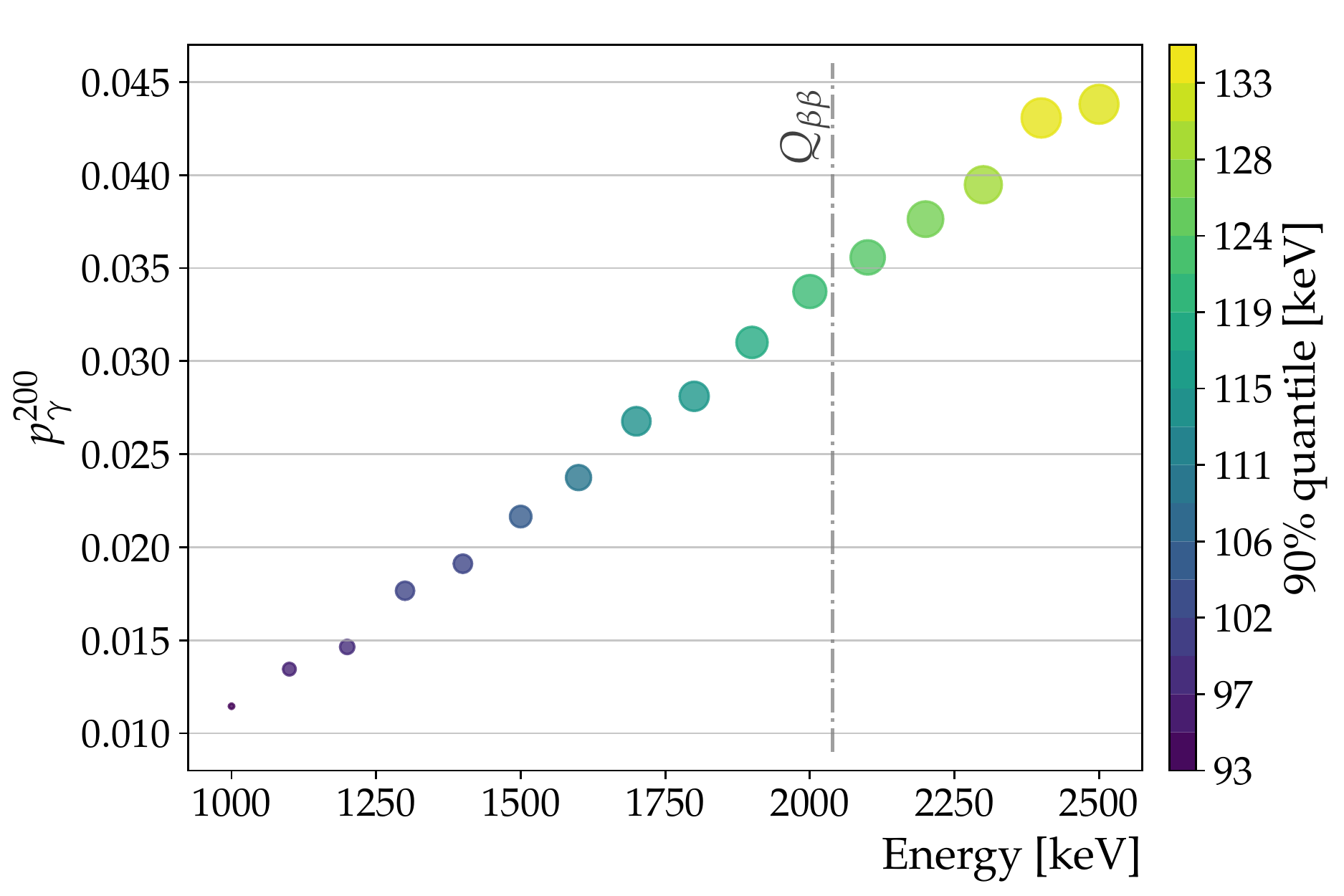}
\caption{Probability of emitting a photon with energy greater than 200 keV, as a function of the energy of the primary electron. Both the marker size and color indicate the 90\% quantile of the gamma-rays' energy spectrum.}
\label{fig:PvsE}       
\end{figure}

To characterize the Bremsstrahlung radiative process, we show in Fig.~\ref{fig:PvsE} the probability of producing photons with energy larger than $200$ keV\footnote{$200$ keV is the energy threshold which corresponds to photon scattering length of $\approx$ $10$ mm in germanium} as a function of the energy of the primary electron. This increases linearly from $1\%$ to $4.5\%$ for kinetic energies from $1$ to $2.5$ MeV. With the help of the color scheme and the marker area, both indicating the $90\%$ quantile of the Bremsstrahlung energy spectrum, Fig.~\ref{fig:PvsE} also shows that, with increasing primary energy, Bremmstrahlung gammas are also more energetic.
This means that, with increasing energy, not only the spatial distributions get on average broader, but the fraction of exceptionally enlarged ones increases, as well. 
This is reflected in \rn\ distributions as a shift of events from the peak to the tail, as shown in Tab.~\ref{tab:r90_char}, under the \emph{peak-to-tail} label.\footnote{This value is calculated taking the ratio of the integral outside and inside the main \rn\ peak. The delimiting point was arbitrarily taken $4\sigma$ away from the centroid of the peak.}

\begin{table}
  \centering
  \begin{tabular}{cccc}
  \hline\noalign{\smallskip}
    Energy  &	\emph{\rn\ peak}	& $\sigma_{R_{90}}$ &	 \emph{peak-to-tail}   \\       
  $[MeV]$	  & $[mm]$	          & $[mm]$            &  $[\%]$	          \\  
  \noalign{\smallskip}\hline\noalign{\smallskip}
  1.0       &	      0.25  	    &       0.07        &	 	4.7	            \\	
  1.5	      &	      0.43  	    &       0.13        &  	5.4	            \\
  2.0       &	      0.63  	    &       0.18        &  	6.5	            \\
  2.5	      &	      0.84	      &       0.24        &  	6.5	            \\
  
  \noalign{\smallskip}\hline
  
  \end{tabular}
  \caption{Summary of the parameters of interest for the description of spatial distributions of electrons in germanium, and their evolution with energy.}
  \label{tab:r90_char}       
\end{table}

\subsection{Event samples of interest for \onbb\ searches}
\label{subsec:onbb-like_events}

\begin{figure}[htb]
  \centering
  \includegraphics[width=0.47\textwidth]{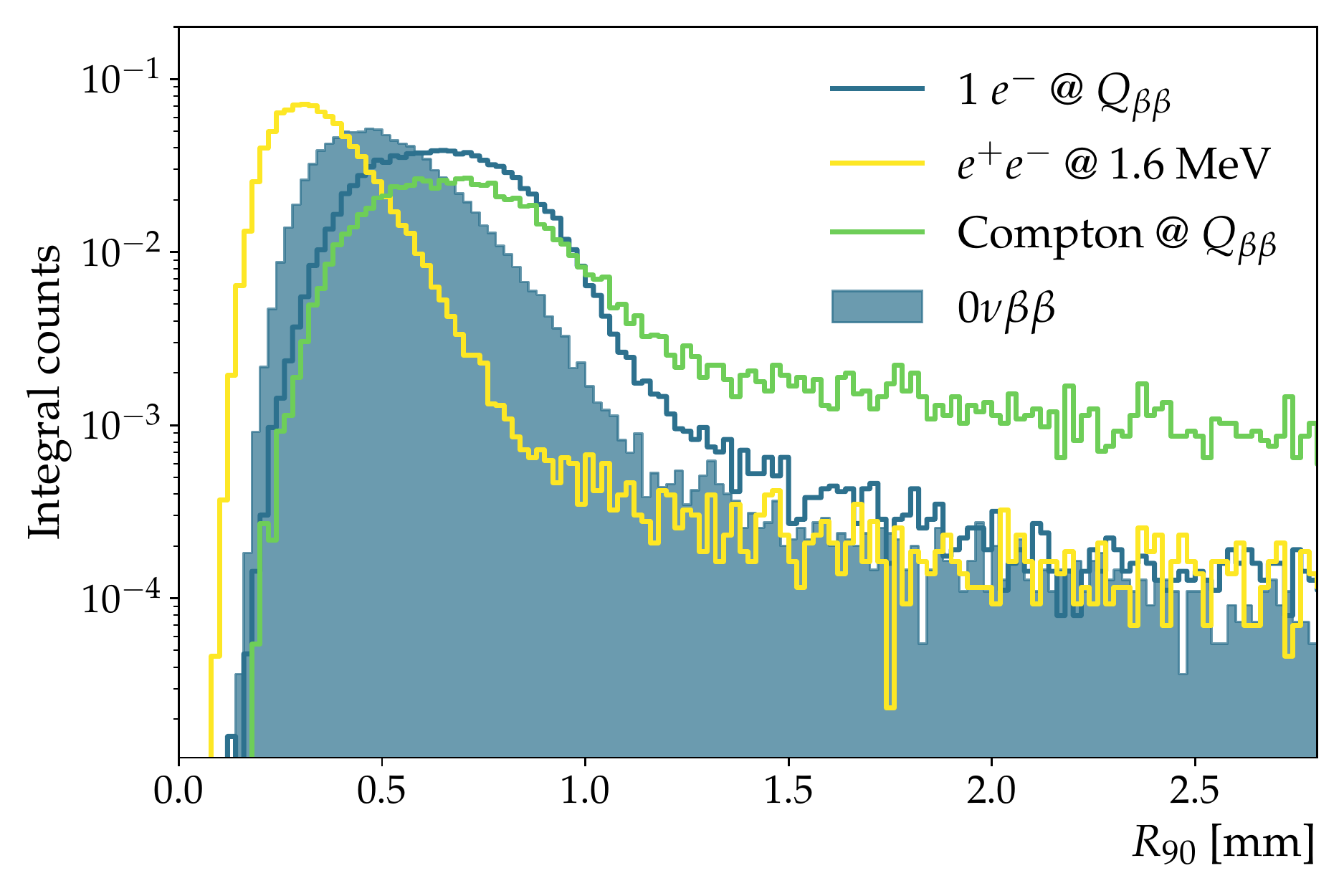}
  \caption{Comparison of the \rn\ distribution from \onbb\ events (filled histogram) with a single electron with initial energy equal to \qbb\ (empty blue histogram). The yellow and green histograms extend the comparison to the samples of interest for \onbb\ searches, namely \ogee\ events at $1.6$ MeV (yellow) and Compton-scattered events around \qbb\ (green). Histograms are normalized on the total number of events in each sample.}
  \label{fig:singlevsdouble}       
\end{figure}

The experimental signature of \onbb\ is an energy deposition of 2 electrons sharing the full $Q$-value of the decay (\qbb=$2039$ keV). They are likely to share evenly the decay energy, though very asymmetric cases are also possible. Their spatial distributions will hence resemble those of single-electron events with energy in the range $1-2$ MeV. Fig.~\ref{fig:singlevsdouble} shows the \rn\ distribution for \onbb\ and single-electron events with initial energy equal to \qbb. The distributions have similar shape, with the Gaussian peaks slightly shifted due to the fact that not one, but two electrons are present in \onbb\ events. 

Every technique aiming to select \onbb\ events must be calibrated with samples with a topology which is similar to \onbb. Samples of \onbb-like events are ty\-pically produced by irradiating Ge detectors with a $^{228}$Th calibration source, which provides a 2.6\,MeV gam\-ma\--ray due to the decay of $^{208}$Tl. The $2.6$ MeV gammas can interact through pair-production, creating an electron and positron that share the whole energy minus the mass energy of the electron-positron pair, thus $1.6$ MeV. 
When the positron thermalizes, it annihilates with an orbiting electron, emitting two secondary gamma-rays. If both escape the detector, then the pair-production events have a topology similar to that of two-electron events with shared energy of $1.6$ MeV. For this reason, as shown in Fig.~\ref{fig:singlevsdouble}, the shape of their \rn\ distribution is similar to that of \onbb\ events. Due to the different energy of the two processes, however, the Gaussian peak of pair-production events is shifted to lower \rn\ values.

A second group of \onbb-like events is created by the 2.6\,MeV gamma-rays scattering only once within the detector. This second samples is composed of single-electron events with energy values between zero and the Comp\-ton edge at $2382$ keV.
This sample, however, contains also multiple Compton-scattered events, which typically account for about 50\% of the sample size. Fig.~\ref{fig:singlevsdouble} shows the \rn\ distribution of events with an energy around \qbb\ from the Compton continuum generated by the $2.6$ MeV gamma-ray of $^{208}$Tl. Its Gaussian peak matches that from \onbb\ events, though multiple Compton-scattered events now create a higher tail at high \rn values. 
On the other side, pair-production events can be ef\-fec\-ti\-ve\-ly selected applying a narrow energy cut around the double-escape peak at $1592$ keV, but will still contain multiple-Compton scattered events at the level of a few percent.  

\onbb\ experiments have been using these samples to calibrate their tagging techniques and evaluate the signal detection efficiency. The difference between calibration samples and the actual \onbb\ signal might lead to biases which are difficult to evaluate and are the focus of the next sections.

\section{\onbb-events identification techniques}
\label{sec:aoe}

\begin{figure*}[ht]
  \centering

  \subfloat[][$A/E$ distribution from a primary electron with an energy of 2 MeV. The line histogram shows the total distribution, and the filled ones represent the subsets of events where an energy $E_\gamma$ higher than 50 and 200 keV is converted into Bremsstrahlung radiation.]{\includegraphics[width=.48\textwidth]{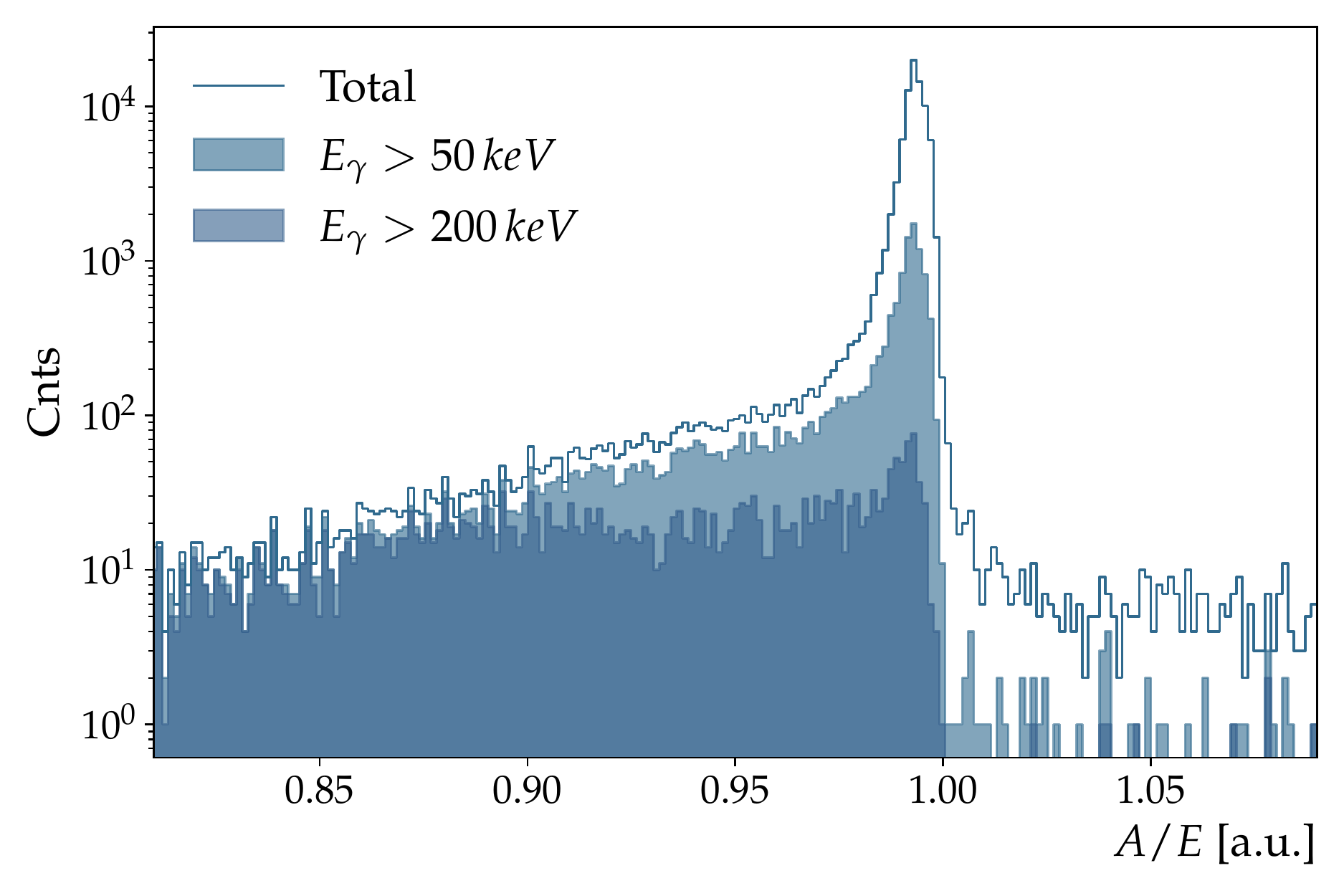}\label{subfig:ha}} \quad
  \subfloat[][Dependence of the $A/E$ distributions on energy. The green line follows the dependence of the peak, the yellow, light and dark green bands the 85\%, 90\% and 95\% quantiles, respectively]{\includegraphics[width=.48\textwidth]{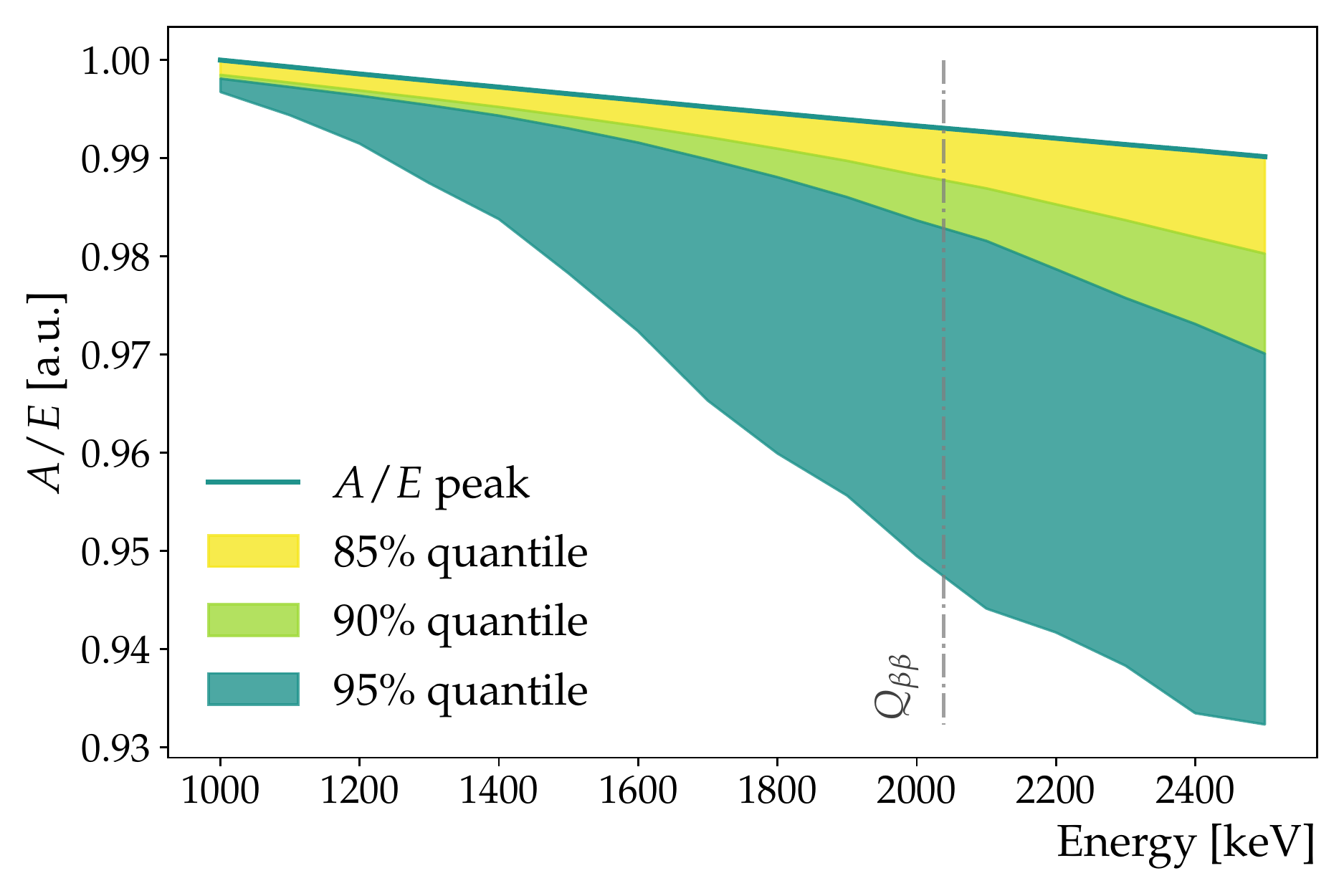}\label{subfig:AoEvsE}} \\

  \caption{$A/E$ distributions from absorption of monoenergetic electrons in germanium, and their dependence on energy.}
  \label{fig:hr90}       
\end{figure*}

\subsection{Signal-background discrimination estimators}
\label{subsec:aoe_single_elec}

Energy depositions in a germanium detector create charge carriers that drift in the detector volume following the electric field lines. Their drift induces a current, which for small anode detectors exhibits a characteristic peak shape when the charge carriers are collected at the anode. As the shape and number of peaks changes according to the event topology, the analysis of its time profile is a powerful tool to discriminate among different event types. 

The $A/E$ technique is a time profile analysis~\cite{DusanBEGe} used by leading \onbb\ experiments~\cite{gerda.PSD.2022, mjd:avse} to discriminate events characterized by multiple interaction sites, which are typical of external gamma background, and keep point-like interactions, typical of the \onbb\ signal. It is based on a single parameter ($A/E$), which is the ratio between the amplitude of the current peak ($A$) and the total deposited energy ($E$). Whenever energy is deposited in multiple sites, the current exhibits several peaks whose height $A_i$ is proportional to the energy $E_i$ deposited in every site. Compared with the case in which the full energy is deposited in a single location, multiple site events exhibit a lower $A/E$ value, as every peak value $A_i$ gets divided by the full energy $E$.

\subsection{Reconstruction of the event topology}
\label{subsec:aoe_reconstruction}

To investigate the relationship between $A/E$ and event topology, the Monte Carlo simulations discussed in the previous section have been post-processed with \texttt{SigGen}~\cite{Radware}, a software package able to simulate the Ge detector physics and the electrical signals expected for each event. Then, the generated signals have been convolved with typical electrostatic noise recorded from an actual setup (resulting in an energy resolution $\sigma_E$ of $1.2$ keV at \qbb), and finally analyzed to calculate the $A/E$ value. 
The $A/E$ distribution are shown in Fig.~\ref{subfig:ha} and \ref{subfig:AoEvsE}. Similarly to the \rn\ distribution, the single-electron events without Bremmstrahlung are reconstructed at similar $A/E$ values creating a Gaussian peak in the distribution. Bremsstrahlung events populate a tail that extends to lower $A/E$ values.
Fig.~\ref{subfig:AoEvsE} shows that the peak in the $A/E$ distribution decreases linearly with energy by about $1\%$ per MeV. The quantiles of the distribution decrease faster and non-linearly with the energy. This feature has been identified during this work for the first time, and its has an important impact on the $A/E$ analysis which will be highlighted in the following sections.

\begin{figure}[tb]
    \includegraphics[width=0.50\textwidth]{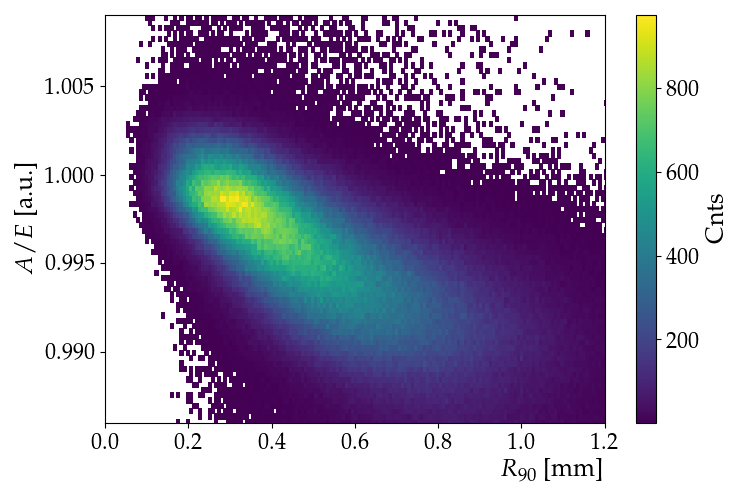}
  \caption{Correlation between $A/E$ and \rn\ for absorption of electrons between 1 and 2.5 MeV.}
  \label{fig:AoEvsR90}       
\end{figure}

The correlation between \rn\ and $A/E$ is explicitly shown in Fig.~\ref{fig:AoEvsR90}, where the two parameters are plotted one against the other for single-electron events with energy between 1 and 2.5 MeV. 
The main feature is the densely populated band with $A/E$ values between $0.99$ and $1$, and \rn\ between $0.2$ mm and $0.8$ mm, representing the correlation between the narrow peaks of \rn\ and $A/E$. As \rn\ gets larger, the energy depositions tend to be less localized, in turn lowering the $A/E$ value. 
Outside the band, we observe events with $A/E > 1$, which are known to be generated by interactions in the volume surrounding the anode \cite{Agostini_SignalModeling}, and Bremsstrahlung events, populating the region of $A/E < 0.99$ and \rn$> 0.3$ mm.
This is consistent with the results discussed in Ref.~\cite{schweisshelm:thesis}.

All small anode detectors will show an inverse proportionality between $A/E$ and \rn. Our results are hence of general interest and apply to a large group of detector geometries. They are however qualitative, as the actual strength of the inverse proportionality depends on the detector geometry and the resulting electric field.

\subsection{Calibration of \onbb-events identification criteria}
\label{subsec:calibration}

The selection criteria used to extract a data sample with enhanced fraction of \onbb\ events consist on an energy-dependent two-sided cut on the $A/E$ parameter, selecting all events in the peak of the distribution and rejecting those in the tails. The cut values are inferred from dedicated calibration measurements creating samples of \onbb-like events. As anticipated in Sec.~\ref{subsec:onbb-like_events}, such events are typically obtained by irradiating Ge detectors with a \Th\ source. The 2.6 MeV gamma-rays from its decay create two classes of \onbb-like events:
\begin{itemize}
  \item \emph{Single Compton Scattering}: events in which the gammas scatter only once in the detector, before leaving the active volume. In such cases the energy between zero and the Compton edge at $2.4$ MeV is transferred to a single electron 
  \item \emph{Pair-production with double escaping gammas} (\ogee): events in which the $2.6$ MeV gammas interact through pair-production inside the detector creating an elec\-tron-positron pair. If the two photons from the annihilation of the positron escape the detector, the energy deposition occurs only through the electron and the positron, which deposit $2614 - 2\cdot511 = 1592$ keV in the detector
\end{itemize}
The first class of events provides \onbb-like samples at different energies, up to the Compton edge, and can be used to study the energy dependence of the $A/E$ parameter highlighted in Fig.~\ref{subfig:AoEvsE}. The standard calibration procedures performs this operation by tracking centroid and width of the $A/E$ peak for single Compton-scattering samples at selected energies between $1$ and $2.4$ MeV. The energy dependence thus obtained is then used as a correction, such that the $A/E$ centroids align on a constant value and exhibit the same width. This standardized approach was conceived without noticing that the centroid and the quantiles in the $A/E$ distributions have different energy dependencies \cite{DusanBEGe, gerda.PSD.2022}. 
On the other hand, \ogee\ events occur at a fixed energy, and the contamination of single and multiple Compton-scattered events can therefore be limited to a few percents by applying a strict energy selection. For this reason, the sample of \ogee\ events is used as main proxy to tune and fix the efficiency of the \onbb-selection.
Histori\-cally, this has been done by setting a lower threshold on their $A/E$ distribution, which keeps $90\%$ of the sample events.

\subsection{Estimation of the \onbb-tagging efficiency}
\label{subsec:onbb-tagging}

The reconstruction of the event topology with the $A/E$ parameter depends on several factors, like the detector geometry and, to a certain extent,  the interaction position. Indeed, it is well-known that small anode detectors feature events with amplified $A/E$ for events occurring in the volume surrounding the anode \cite{Agostini_SignalModeling}. Therefore, $A/E$ distributions have features which not only depend on the starting kinetic energy, but also on the event sample. A detailed modeling of the calibration and \onbb\ samples is thus mandatory for a precise estimation of the tagging efficiency of \onbb-events.

\begin{figure}[htb]
  \centering
  \includegraphics[width=0.48\textwidth]{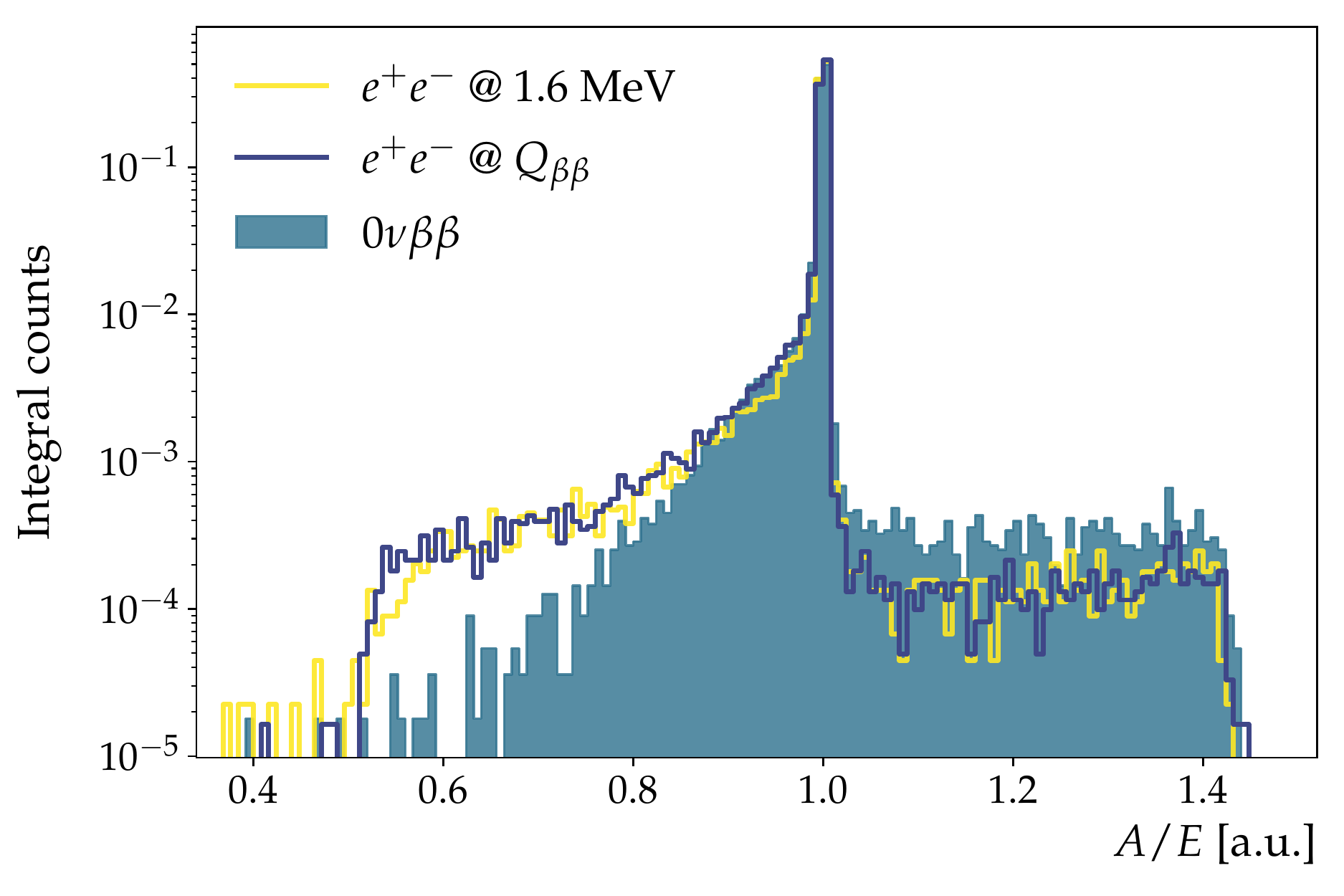}
  \caption{$A/E$ distributions of pair production events and \onbb.}
  \label{fig:DEPvs0nbb}       
\end{figure}

The fact that the standard calibration procedure tracks the $A/E$ distribution peak centroid (and not the desired quantile) results in an energy-decreasing tagging efficiency for \onbb-like events.
To estimate the bias at \qbb, we simulated \ogee\ events occurring exactly at \qbb\ and followed the standard calibration procedure with \Th. 
Fixing the tagging efficiency of \ogee\ events at $1.6$ MeV to $90\%$ yields an efficiency of $(86.5\pm0.4)$\% at \qbb. Fig.~\ref{fig:DEPvs0nbb} shows the $A/E$ distributions of the two \ogee\ samples, after the standard correction of the energy dependence. As an effect of the calibration procedure, the two centroids are centered around the same value (which is arbitrarily set to $1$). At $A/E$ values around $0.95$, the distribution of \ogee\ at \qbb\ shows an excess of events. As seen in Sec.~\ref{subsec:aoe_reconstruction}, this is due to a higher production of secondary Bremsstrahlung gamma-rays, which increases the background-like character of the sample and therefore decreases its overall tagging efficiency.

Moreover, calibrating the \onbb-tagging efficiency on a different sample brings an additional bias. Indeed, being $^{76}$Ge homogeneously distributed, so will its decay products, while \ogee\ events are more likely to occur on detector lateral surfaces and corners, where the probability for the annihilation photons to escape detection is maximal. To evaluate the impact of this geometrical difference, we used a Monte Carlo sample of \onbb-events and followed the standard calibration procedure with \Th. This yields a tagging efficiency of $(85.2\pm0.4)\%$\footnote{The value here shows some tension with what we reported in \cite{comellato:modelingIC}, as it has been obtained with slightly different analysis routines.}, i.e. a further $1\%$ reduction.
Fig.~\ref{fig:DEPvs0nbb} shows also the $A/E$ distribution of the simulated \onbb\ events in comparison to \ogee\ at 1.6 MeV and at \qbb. It shows that this further reduction is a complex balance of features in the whole $A/E$ spectrum. First, \ogee\ events exhibit a higher tail at very low $A/E$ values. Being present for \ogee\ events at 1.6 MeV as well, it suggests that its origin is intrinsically related to the dynamics of pair-creation. On the high $A/E$ side, being \onbb s homogeneously distributed, they probe the volume region surrounding the anode more significantly, which gives signals with amplified $A/E$. 

\section{Energy dependent event discrimination with a \cofs\ source}
\label{sec:energy-dependent-tagging}

In the previous sections, we have seen that A/E distributions shift and change shape with energy. Also, we have seen that the samples used in the standard calibration procedures, i.e. \ogee\ and Compton-scattered events, generate A/E distributions which actually differ from that expected from \onbb. 
Before the present work, these facts were, to some extent, already known. Pulse Shape Simulations (PSS) were used in the state-of-the-art experiments to evaluate the systematic uncertainties due to these differences \cite{gerda.PSD.2022}. 
With a thorough validation, the values from PSS could eventually be used as a central value for the \onbb-tagging efficiency. 
However this has been avoided so far as tens of even hundreds detectors of different shapes are operated in the setup, making the tuning and validation of their individual modeling extremely challenging. 
In this section we study the option of a data-driven calibration of the energy dependence avoiding the bias discussed in the previous section. As it's the most significant contribution to the overestimation of the \onbb-tagging efficiency with a value of 4\% out of a total of 5\%, this would reduce the systematic uncertainties to 1\%, without requiring a detector-by-detector modeling. This could be achieved exploiting the decays of a \cofs\ source.

The \cofs\ source used in this work has been custom-produced by the Jagiellonian University in Krakow and its results have been cross-checked with post-processed Monte Carlo simulations.
Its energy spectrum is characterized by several high-energy gamma-lines up to $3.6$ MeV. For this reason, it was early recognized as a valuable source to calibrate germanium detectors \cite{barker1967, gehrke1971, proc:co56}. In our work, we considered five gamma-rays with energy higher than 2.5 MeV and branching ratio higher than 1\%, as the probability of creating electron-positron pairs for them is high enough to give samples of \ogee\ events with a small background contamination. 
The energies of the considered \ogee\ samples are listed in Tab.~\ref{tab:SP} next to their parent gamma-rays. Lying in the range between 1.5 and 2.5 MeV, they offer the opportunity to investigate the energy dependent tagging efficiency of \onbb-like events with many data points and additionally to infer its value directly at \qbb. 

\begin{table}
  \begin{tabular}{ccccl|cl}
  \hline\noalign{\smallskip}
      &			&		& \multicolumn{2}{c}{\textbf{Data}} & \multicolumn{2}{c}{\textbf{Simulations}} \\       
  E (FEP)	& 	E (DEP) 	&	BR	& \multicolumn{2}{c}{SP} & \multicolumn{2}{c}{SP}   \\
  $[keV]$	& $[keV]$	&	\%	& \multicolumn{2}{c}{\%} & \multicolumn{2}{c}{\%} \\  
  \noalign{\smallskip}\hline\noalign{\smallskip}
  2598.5	&	1576.5	&	17	&	88.8	&	(5)	 & 88.6 &	(4)   \\	
  3009.6	&	1987.6	& 	1	&	86.0	&	(30) & 87.4	&	(27)  \\
  3202.0	&	2180.0	& 	3	&	84.6	&	(11) & 85.9 &	(10)	\\
  3253.5	&	2231.5	& 	8	&	85.6	&	(6)	 & 85.0	&	(5)   \\
  3273.1	&	2251.1	& 	2	&	83.9	&	(18) & 85.4	&	(16)	\\
  \noalign{\smallskip}\hline
  \end{tabular}
  \caption{Energies of the Double Escape Peaks (DEPs) used in our \cofs~analysis and respective Survival Probability (SP) after $A/E$  cut. Informations on the energy and Branching Ratio (BR) of the relative Full Energy Peaks (FEPs) are also given as reference.}
  \label{tab:SP}       
\end{table}

\begin{figure}[b]
	\centering
	\includegraphics[width=0.35\textwidth]{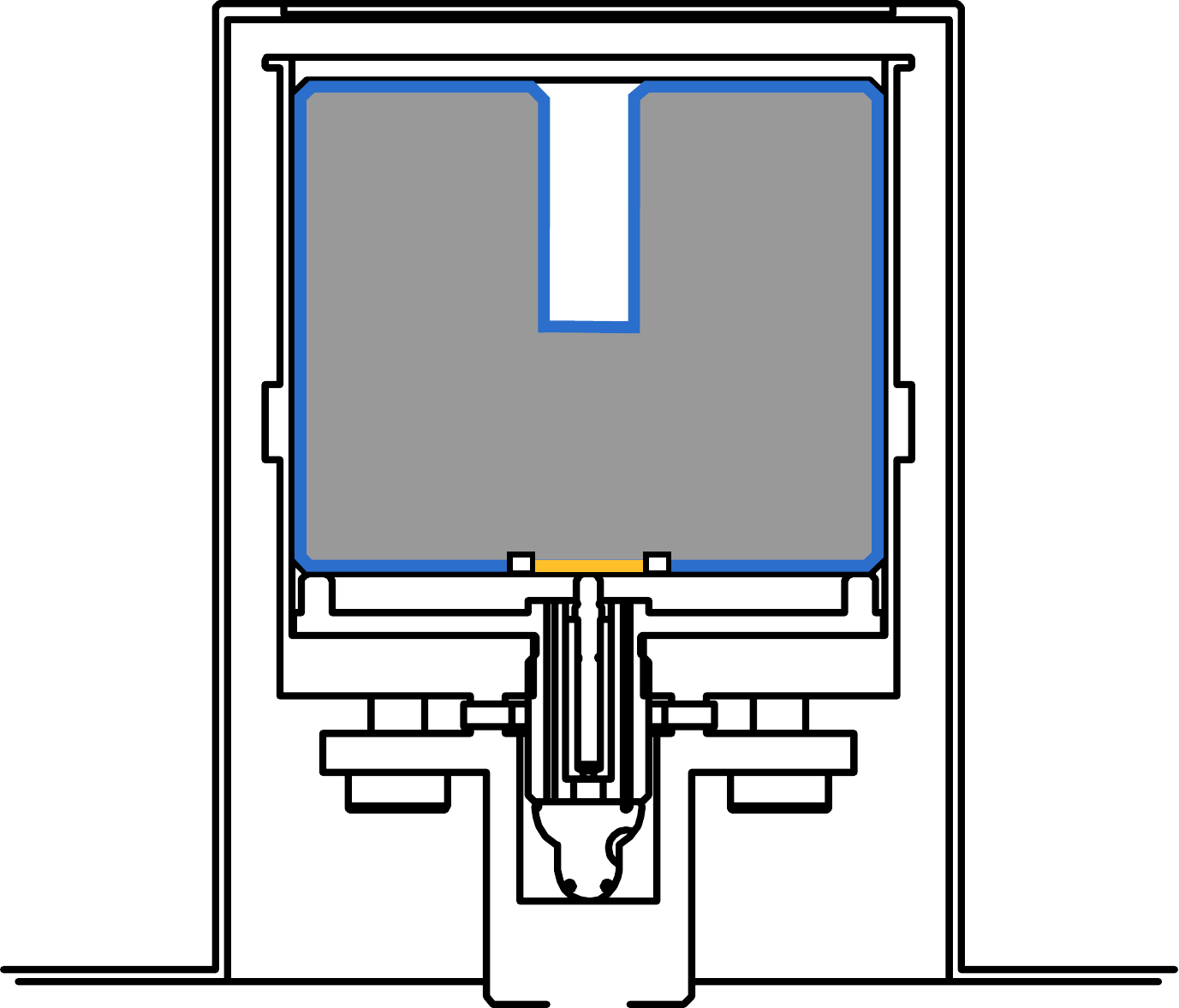}
	\caption{Drawing of the inverted coaxial detector used for the experimental data with the \cofs\ source, inside the vacuum cryostat provided by the vendor. Marked in yellow is the Boron-implanted p$^+$ electrode and in blue the Lithium-diffused n$^+$.}
	\label{fig:gwd6022-drawing}       
\end{figure}

\begin{figure}[t]
  \centering
  \includegraphics[width=0.48\textwidth]{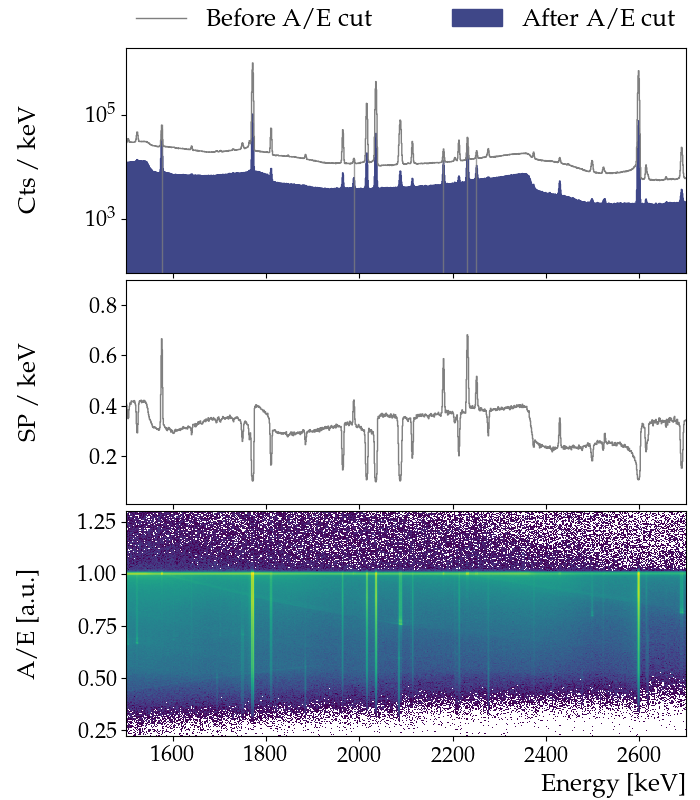}
  \caption{Region of interest of the energy spectrum (top) obtained from a \cofs\ source. The middle panel shows the Survival Probability (SP) of every energy bin after an $A/E$ cut calibrated on a \Th\ source. The bottom panel illustrates how the events in this energy range are distributed in the $A/E$ space. The position of the \ogee\ events in the energy spectrum is marked with a solid gray line.}
  \label{fig:co56_spectrum_aoe}       
\end{figure}

The detector used to record data with this source is shown in Fig.~\ref{fig:gwd6022-drawing}. It is a $1.6$ kg inverted coaxial detector, which has been operated inside its vendor's vacuum cryostat (thus approaching the temperature of $77$ K) and whose geometry is the one used as a reference for the simulations of the present work. More details on the detector and on the validation of the simulations can be found on our previous work \cite{comellato:modelingIC}. The source, which had an activity of $90$ kBq at the time of data taking, was placed $20$ cm away from the cryostat, and the resulting energy spectrum is shown in the top panel of Fig.~\ref{fig:co56_spectrum_aoe}, in the range where the \ogee\ events occur. The empty grey and filled blue histograms show the spectrum respectively before and after the $A/E$ cut calibrated on \Th. The ratio between the two is shown in the middle panel, where the $6$ peaks from \ogee\ events clearly arise from the Compton continuum. The bottom panel shows how all the events from the decays of \cofs\ populate the $A/E$ spectrum. Its description is analogous to Fig.~\ref{fig:AoEvsR90}: the high density horizontal band centered around $1$ represents all the energy depositions which occurr in a single location, while the band below contains all the events which deposit energy in more than one site, and the region where $A/E$ is higher than $1$ represents those which occur in the volume surrounding the anode. In this spectrum, \ogee\ and single Comp\-ton events are distributed around $A/E=1$, while multiple Compton scatterings populate the region where $A/E<1$.

\begin{figure}[t]
  \centering
  \includegraphics[width=0.48\textwidth]{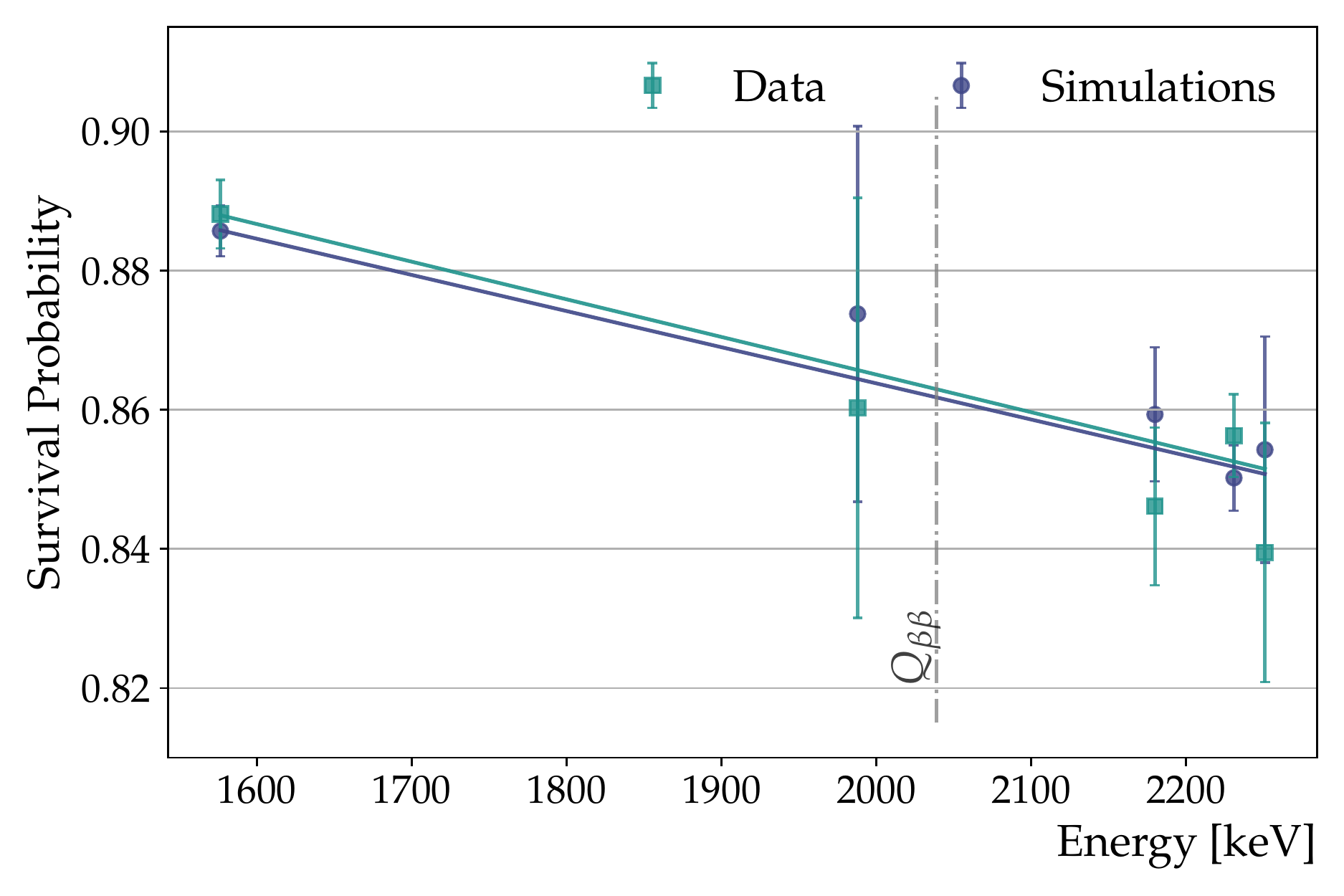}
  \caption{Survival probability of \ogee events from \cofs~after $A/E$ cut. Light blue squares indicate results from simulated data, dark blue circles from simulations.}
  \label{fig:SPvsE}       
\end{figure}

The \ogee\ tagging efficiency is extracted for experimental (simulated) data following the standard calibration procedure with \Th\ described in Sec.~\ref{subsec:calibration}. This yields the results listed in the fourth (fifth) columns of Tab.~\ref{tab:SP}, and shown with the light blue squares (dark blue circles) in Fig.~\ref{fig:SPvsE}. 
The agreement between simulated and experimental data is within the statistical uncertainties, which corroborates the results of our simulation campaigns. 
In both datasets, the fraction of tagged events is systematically lower than 90\%, even at $1576$ keV. The tagging efficiency for this peak is $1\%$ smaller than the value at $1592$ keV from the calibration on \Th. This comes from the complexity of the \cofs\ spectrum (which exhibits many features in the $A/E$ space, see Fig.~\ref{fig:co56_spectrum_aoe}), which prevents the analysis from selecting a completely pure sample of \ogee\ events. For this reason, we take $1\%$ as systematic uncertainty of the method.
Despite the small offset, the tagging efficiency decreases as a function of energy. A linear interpolation of these results yields a value of $(86.3\pm0.4)\%$ at \qbb\ for experimental data and $(86.2\pm0.3)\%$ for simulations, which is in good agreement with the results on the \ogee\ events directly at \qbb\ of Sec.~\ref{subsec:calibration}.


\section{Discussion and conclusions}
\label{sec:conclusion}

In this paper we characterized the spatial distributions of electron energy depositions in germanium and studied how they are affected by the initial kinetic energy of the electron. 
We identified a connection between the spatial distributions and the estimator used in \onbb\ experiments to discriminate single and multiple interaction sites, i.e. the $A/E$ parameter.
We used this information to review the standard calibration procedures with a \Th\ source. 
In particular, we quantified that the standard calibration procedures used so far by experiments result in two biases in the determination of the \onbb-tagging efficiency. The first is an energy-dependent overestimation of the efficiency at the level of a few percents (4\% for our simulated geometry). The second is a 1\% contribution due to using a different event sample than \onbb.
Additionally, we estimated the effect of different electronics noise levels on the discrimination efficiency of \onbb\ events. A total reduction of 3\% is obtained when an electronics noise at the level of the state-of-the-art experiments (resulting in an energy resolution $\sigma_E$ of $1.5$ keV at \qbb) is used\footnote{Though an increase in the \onbb-tagging efficiency is desirable, this comes at the expense of a higher increase in the acceptance of background, as also stated in \cite{comellato:modelingIC}}. In future experiments, which aims at intermediate noise levels between that of a single detector and of present large-scale experiments, a \onbb-tagging efficiency within $85\%$ and $87\%$ could therefore be foreseen.

Though these biases are taken into account in the systematics uncertainties of the state-of-the-art experiment, we showed that, with a \cofs\ source, these can be strongly reduced from 5\% to 1\%.
Furthermore, as \cofs\ provides samples of almost pure \onbb-like events in a broad energy range, new calibration routines could be developed for the correction of the $A/E$ energy dependence, resulting in a bias-free tagging efficiency of signal-like events at all energies.
As three out of five \ogee\ events of \cofs\ come from very low branching ratio gamma-rays, a calibration with sufficient statistics to observe the $2-4\%$ reduction in the tagging efficiency requires very long acquisition times. For future-generation experiments, which are scheduled to acquire data for a decade, the required counting statistics could be gained with multiple or periodical calibration campaigns.
Given \cofs's half-life of only $77.3$ days, this would require a regular production of radiation sources and a calibration plan optimized for this application, which is beyond the scope of this work.

\begin{acknowledgements}
We are very grateful to Grzegorz Zuzel and the Jagiellonian University in Krakow for providing the valuable \cofs\ source for this work. We are also thankful to all the members of the \gerda\ and \legend\ collaborations, and in particular to David Radford, for their valuable feedback. This work has been supported in part by the European Research Council (ERC) under the European Union’s Horizon 2020 research and innovation programme (Grant agreement No. 786430 - GemX), and by the Science and Technology Facilities Council, part of U.K. Research and Innovation (Grant No. ST/T004169/1). M.A. acknowledges also the support of the UCL Cosmoparticle Initiative.
\end{acknowledgements}


\printbibliography[heading=bibintoc]

\end{document}